\documentclass[aip,rsi,reprint,graphicx]{revtex4-1} % for checking your page length
\usepackage{graphicx}% Include figure files
\usepackage{dcolumn}% Align table columns on decimal point
\usepackage{bm}% bold math
\usepackage[mathlines]{lineno}% Enable numbering of text and display math
\usepackage{amsmath}
\begin{document}

%\preprint{AIP/123-QED}

\title[]{Measurement of incandescent microparticle acceleration using stereoscopic imaging}
%\title[]{Measurement of high-temperature microparticle acceleration through imaging}
\thanks{Contributed paper published as part of the Proceedings of the 22nd Topical Conference on High-Temperature Plasma Diagnostics, San Diego, California, April, 2018.\\}
%{Analysis of the Forces on Particles Generated by Exploding-Wire}
%\thanks{Footnote to title of article.}

\author{Pinghan Chu
}\email{pchu@lanl.gov.}
\author{Bradley T. Wolfe}
\author{Zhehui Wang}\email{zwang@lanl.gov.}
\affiliation{ 
Los Alamos National Laboratory, Los Alamos, New Mexico 87545, USA
}%

\date{\today}
\begin{abstract}
Microparticles ranging from sub-microns to millimeter in size are a common form of matter in magnetic fusion environment, and they are highly mobile due to their small mass. Different forces in addition to gravity can affect their motion both inside and outside the plasmas. Several recent advances open up new diagnostic possibilities to characterize the particle motion and their forces: high-speed imaging camera technology, microparticle injection techniques developed for fusion, and image processing software. Extending our earlier work on high-temperature 4D microparticle tracking using exploding wires~\cite{Wang:2016}, we report latest results on time-resolved microparticle acceleration measurement. New particle tracking algorithm is found to be effective in particle tracking even when there are a large number of particles close to each other. Epipolar constraint is used for track-pairing from two-camera views. Error field based on epi-geometry model is characterized based on a large set of 2D track data and 3D track reconstructions. Accelerations based on individual reconstructed 3D tracks are obtained. Force sensitivity on the order of ten gravitational acceleration has achieved. High-speed imaging is a useful diagnostic tool for microparticle physics, computer model validation and mass injection technology development for magnetic fusion.
\end{abstract}

%\pacs{}% PACS, the Physics and Astronomy
                             % Classification Scheme.
%Use showkeys class option if keyword
                              %display desired
\maketitle
\section{Introduction}
\label{sec:introduction}
Production of microparticles or dust of different sizes is not avoidable in burning plasmas like ITER~\cite{Federici:2001,Wang:2008}. Due to their high mobility, one of the main concerns is that the microparticles could migrate inside the closed magnetic flux surfaces and contaminate the plasma core. For low-Z atoms like hydrogen to carbon, the undesired effect is fuel density dilution. For high-Z atoms like tungsten, radiative cooling could be detrimental~\cite{Cecchi:1980}. In ITER-like plasmas, beryllium as an impurity cannot exceed 15 mg in the plasma core. Tungsten as an impurity cannot exceed 0.6 mg in the core. The dynamics of microparticles can be complex due to the interplay of many forces associated with plasma flow, edge pressure and density gradient, edge turbulence, magnetic field strength, heating and ablation. Gravity in comparison was shown to be relatively small in fusion-edge-like plasma conditions~\cite{Wang:2007}. Additional concerns with the microparticles include the spread of tritium and neutron-activated radioactive materials throughout the vacuum chamber, exacerbating the tritium retention and removal problems. Experimental techniques such as high-speed imaging sometimes is the only experimental option to examine the microparticle dynamics. Furthermore, imaging of high-temperature particles or dust can contribute to the understanding of plasma-dust interactions and plasma-surface interactions and development of first wall materials~\cite{Wang:2008b}.

Several recent advances open up new diagnostic possibilities to characterize the particle motion and their forces: high-speed imaging camera technology, microparticle injection techniques developed for fusion, and image processing software. About 10 years ago~\cite{Wang:2007}, commercial fast cameras can only take a few frames of images at a rate above 10$^6$ frames per second (`mega-frame cameras') and with a relatively limited field of view. In comparison, mega-frame cameras can now capture movies continueously for seconds at a time~\cite{Wang:2016}. A growing number of mass injection techniques are used in magnetic fusion~\cite{Wang:2016b}, allowing controlled mass injection and therefore quantitative analysis of microparticle dynamics. Additionally, the broad uses of imaging techniques, including applications in machine vision, have led to rather sophisticated tool kits that are availability for three-dimensional (3D) particle tracking from multiple camera views~\cite{Longuet-Higgins:1981,Hartley:1992}.

In 2016, an exploding-wire apparatus was built to generate high-temperature molten metallic microparticles, emulating the microparticles anticipated near the plasma-facing surfaces in magnetic fusion~\cite{Wang:2016}. The motion of these microparticles was then recorded by two fast cameras in-sync. We showed that time-resolved 3D motion, or `four-dimensional' (4D) tracking of high-temperature microparticles was feasible. These cameras took images at a frame rate of $10^5$ Hz, {\it i.e.}, the inter-frame time of $10^{-5}$ sec and an exposure time of a few $\mu$s.  Metallic microparticles with sizes ranging from 10 to 50 $\mu m$ were generated along the wire after an initial high-voltage (up to 2 keV) high-current (estimated to be a few kA) pulse. High-temperature molten microparticles were created within 10 to 20 $\mu$s based on particle incandescence and camera images.
Two algorithms showed similar results for 4D reconstruction of microparticle motion without statistical analysis. However, it showed the difficulty of successfully pairing a large number of tracks in the left and right cameras. 

In this work, the combination of 2D track identification using nearest velocity with track pairing using both epipolar constraint and temporal brightness are used in the analysis of a large number of tracks. It requires additional process to interpret 3D image data from two fast cameras in-sync. This method provides a robust synthetic diagnostics for understanding of plasma-dust interactions. This algorithm is shown to be effective in particle track identification; here we considered six data sets using the same calibration (Shot118, Shot119, Shot120, Shot216, Shot229, and Shot232). First we will summarize the calibration procedure by considering thousands of matched position data pairs together in Section~\ref{sec:calibration}. Then we will describe the new  algorithm of the 3D tracking using new constraints in Section~\ref{sec:3Dtracking}. In the end, we will present the result of the force analysis on these 3D tracks in  Section~\ref{sec:force}. Theoretical models based on the force analysis will be presented elsewhere.

\section{Error field analysis}
\label{sec:calibration}
When using a single camera and conventional optics for imaging, information such as the depth of the object is lost. When using more than one cameras, similar to human beings' two eyes, we can derive the depth information and reconstruct the three-dimensional (3D) coordinates of objects. This is so-called {\it triangulation} in computer vision. Figure~1 of Ref.~\cite{Wang:2016} showed a schematic of a so-called epipolar geometry using two cameras. A point $M$ in 3D is projected to two points $m_l = (u_{l},v_{l})$ and $m_r=(u_{r},v_{r})$. $m_l$ is in the image plane L of the left camera, and $m_r$ the image plane R of the right camera, respectively. $C_l$ is the optical center of the left camera while $C_r$ is the optical center of the right camera. The triangle $MC_lC_r$ forms a epipolar plane. All points on the line $\overline{MC_l}$ in 3D are projected to $m_l$ in the image plane L. However, the points on the line $\overline{MC_l}$ are projected to different points in the image plane R. Only one point $m_r$ corresponds to the 3D point $M$. This is called {\it epipolar constraint}.
The two points, $m_l$ and $m_r$, can be correlated through the epipolar constraint as
\begin{align}
    \tilde{m}_r^T F \tilde{m}_l = 0,
    \label{eq:fundamentalmatrix}
\end{align}
where $F$ is the fundamental matrix~\cite{Longuet-Higgins:1981,Hartley:1992}, $\tilde{m}_{l}=(u_{l},v_{l},1)^T$ and $\tilde{m}_{r}=(u_{r},v_{r},1)^T$. Given 8 corresponding or more point pairs in the image of the left and right cameras, the fundamental matrix can be computed
by solving a set of linear equations, so-called {\it eight point algorithm}~\cite{Hartley:1997}. For a point in the image of the right camera, the corresponding epipolar line in the image of the left camera is equal to $F \tilde{m}_{e}$ where $\tilde{m}_{e} = (x,y,1)^T$ and $(x,y)$ are coordinates in the image of the left camera, and vice versa for the data points in the image of the left camera.

\begin{figure}[ht]
\includegraphics[width=0.23\textwidth,height=0.13\textheight]{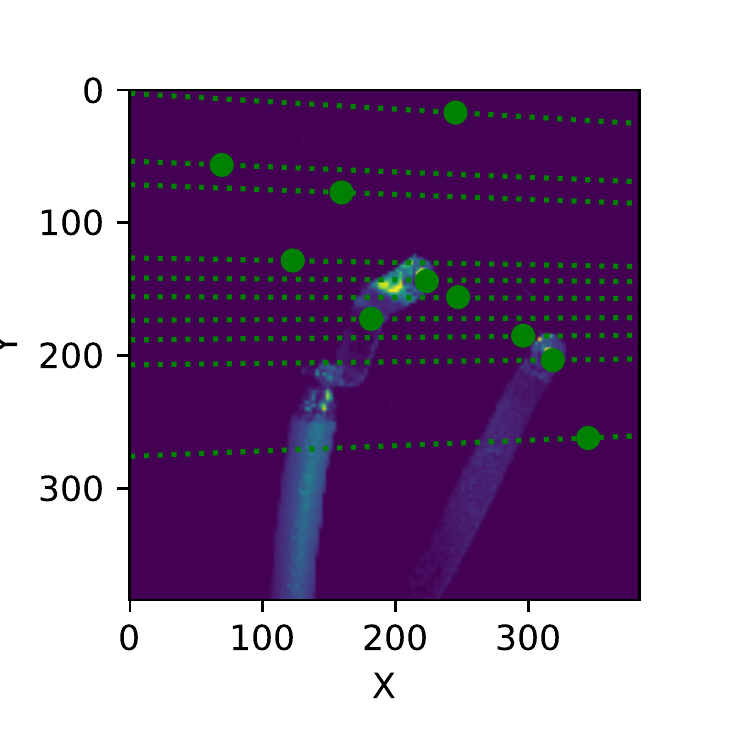}
\includegraphics[width=0.23\textwidth,height=0.13\textheight]{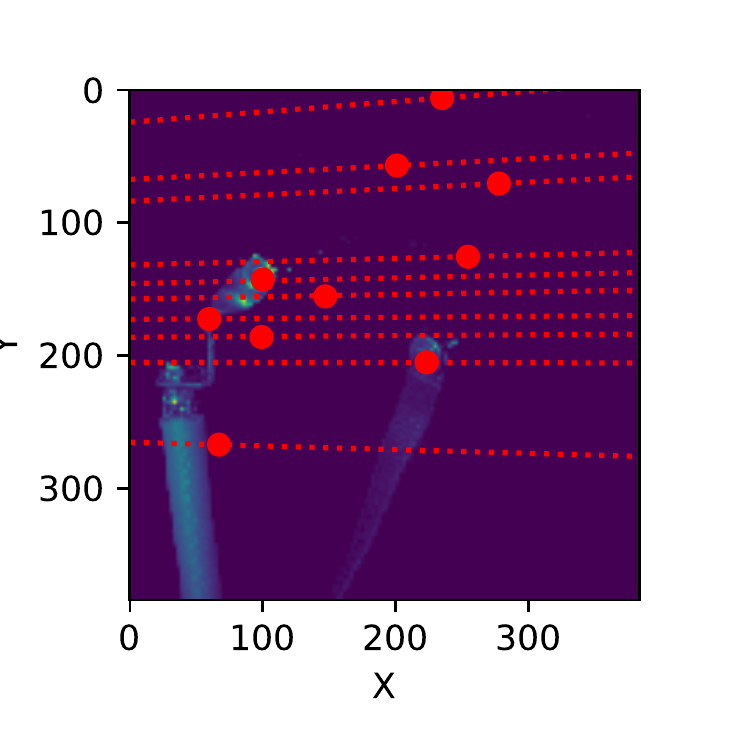}
\includegraphics[width=0.23\textwidth,height=0.13\textheight]{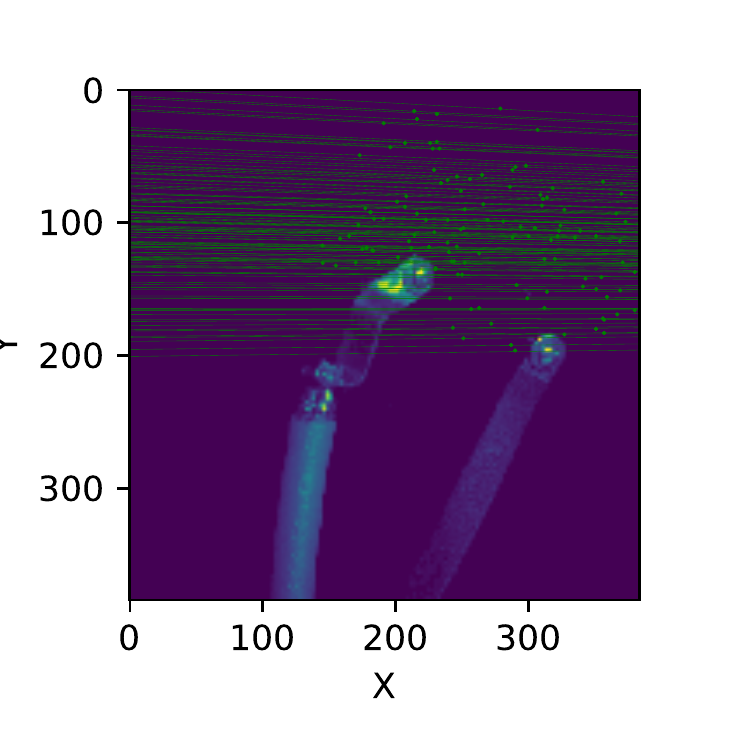}
\includegraphics[width=0.23\textwidth,height=0.13\textheight]{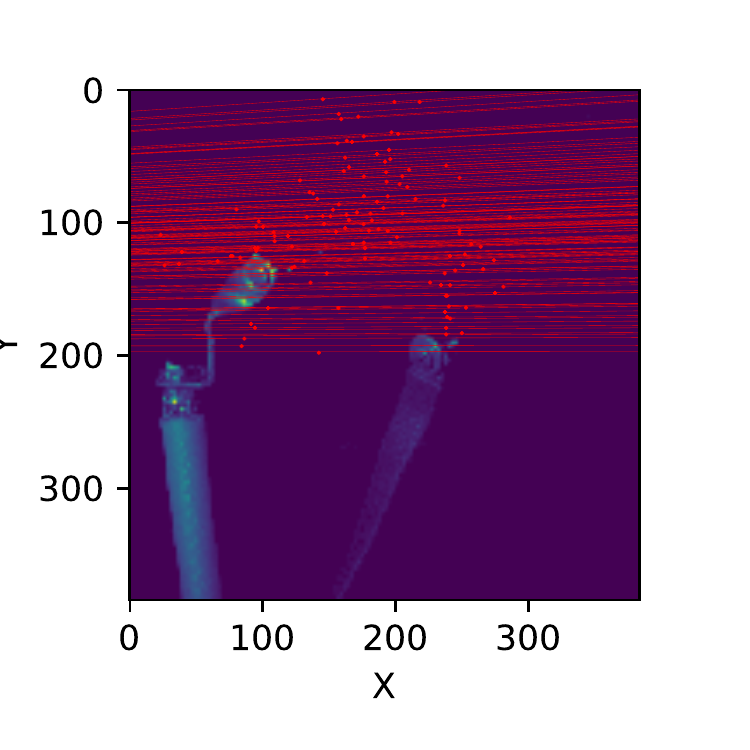}
\includegraphics[width=0.23\textwidth,height=0.13\textheight]{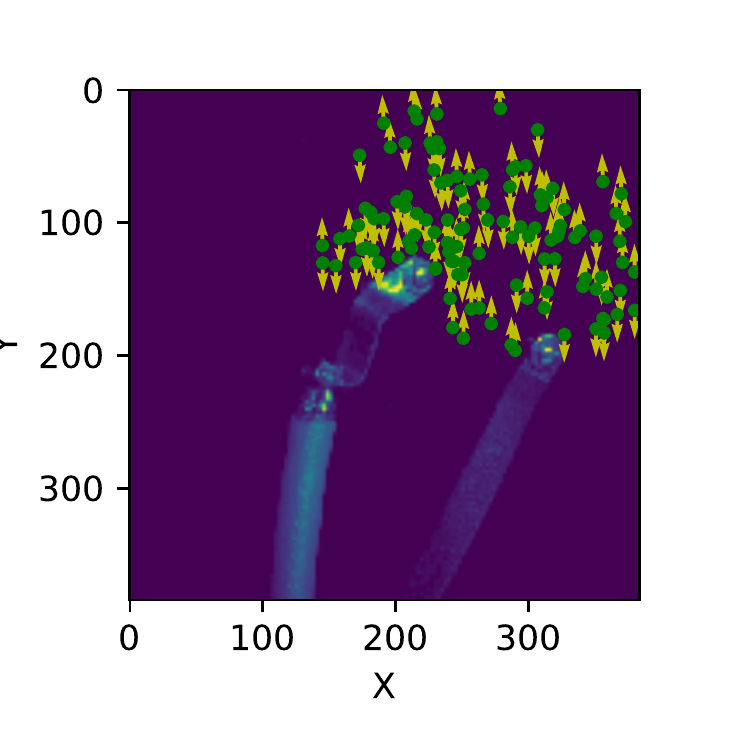}
\includegraphics[width=0.23\textwidth,height=0.13\textheight]{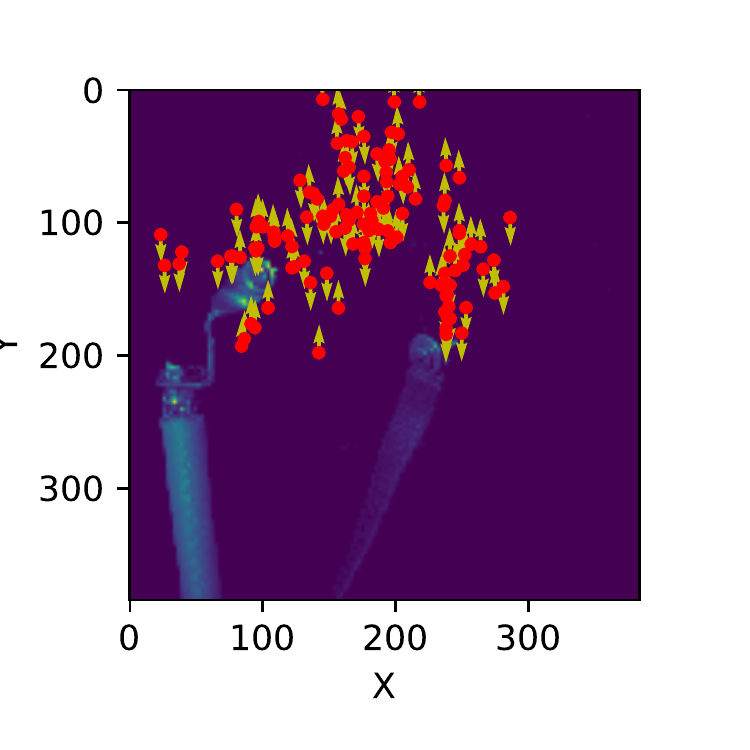}
\caption{Top)Calibration using 10 points. The left image is taken from the left camera and the right image is taken from the right camera. Middle)Calibration of roughly 1000 pairs. Only 10\% of particles are shown here. Bottom)Error field population. Only 10\% of particles are shown here. }
\label{fig:calibration}
\end{figure}

Figure~\ref{fig:calibration} shows the calibration using 10 pairs of matched points in the image of the left and right cameras, which were identified in the study of Ref.~\cite{Wang:2016}, and the corresponding epipolar lines from their pair points in the image of the other camera. These point pairs are manually identified by locating similar points of the objects in two cameras such as electrodes. We want to understand if we can calibrate without a calibration board in a plasma environment. Using the same fundamental matrix, we can search for possible matched pairs in the matched images of the left and right cameras. In principle, a pair of points need to satisfy the epipolar constraint. However, due to the image noise, limited spatial resolution, and other factors, the calibration is not perfect and a point may not fall on its corresponding epipolar line exactly. If so, the minimum distance from the point to its epipolar line, called ``epipolar distance", may be calculated. To match a point in one view with many possible points in the other view, the data points with the smallest epipolar distance are possible pairing candidates. The same procedure has also been applied for data points in the reversed direction. Figure~\ref{fig:calibration} shows the result using calibration of roughly 1000 pairs found in data. 

\begin{figure}[ht]
\includegraphics[width=0.23\textwidth,height=0.15\textheight]{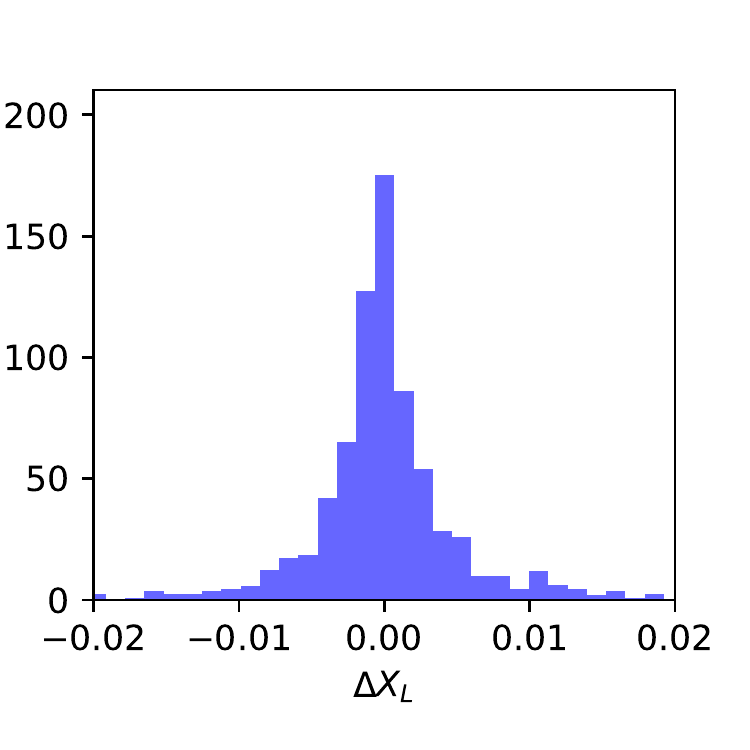}
\includegraphics[width=0.23\textwidth,height=0.15\textheight]{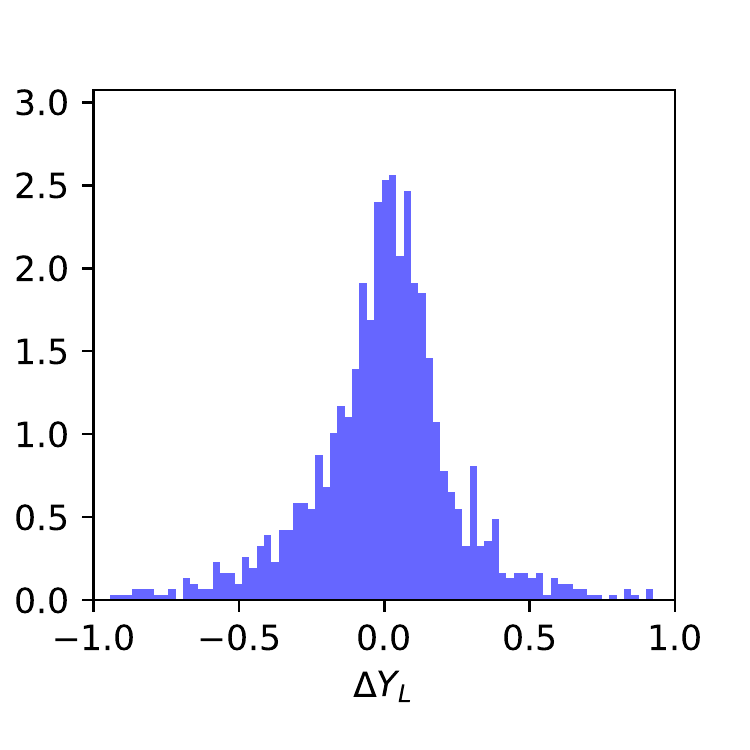}
\includegraphics[width=0.23\textwidth,height=0.15\textheight]{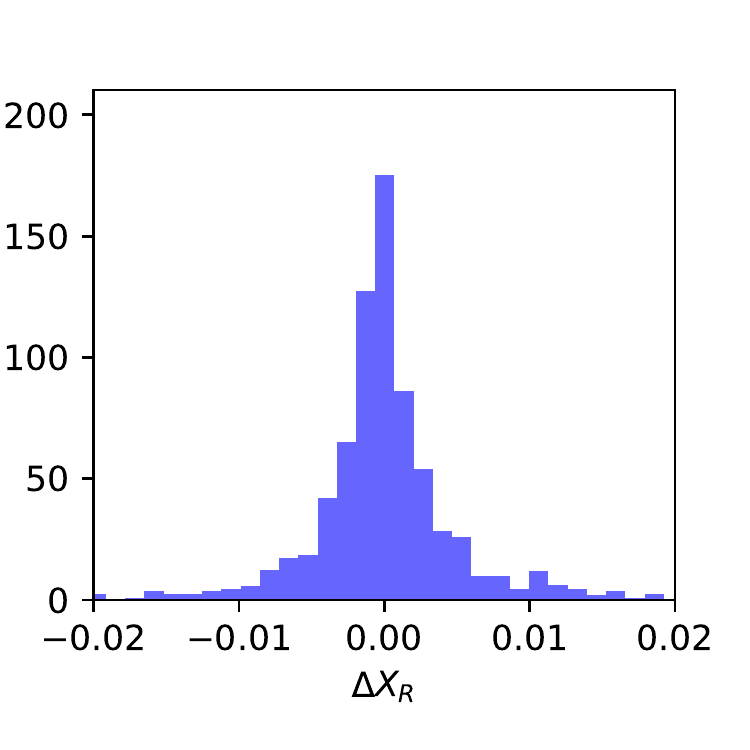}
\includegraphics[width=0.23\textwidth,height=0.15\textheight]{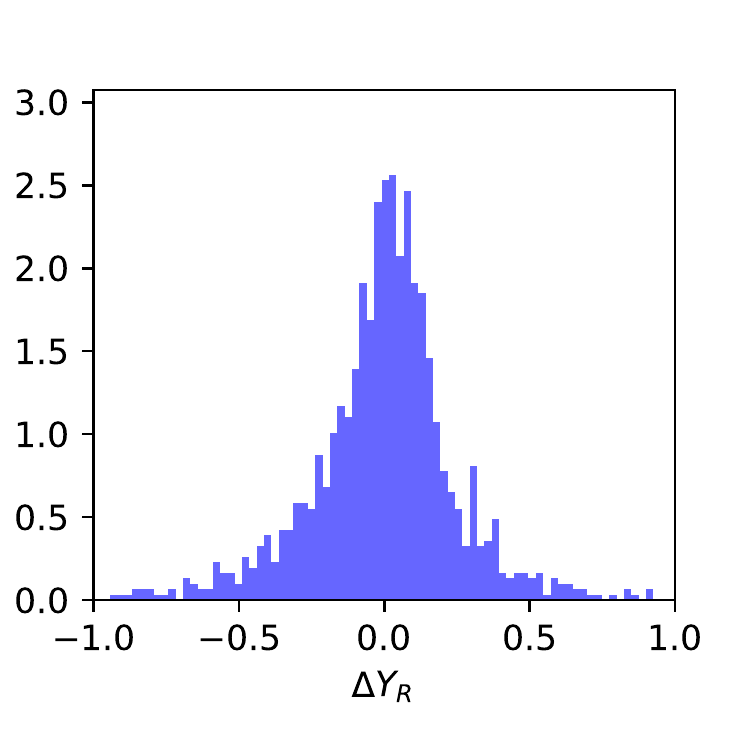}
\caption{Error field population. The deviation on the $x$-axis is about within 0.01 pixel and on the $y$-axis is about within 0.5 pixel.}
\label{fig:errorstat}
\end{figure}

\begin{figure}[ht]
\includegraphics[width=0.21\textwidth,height=0.13\textheight]{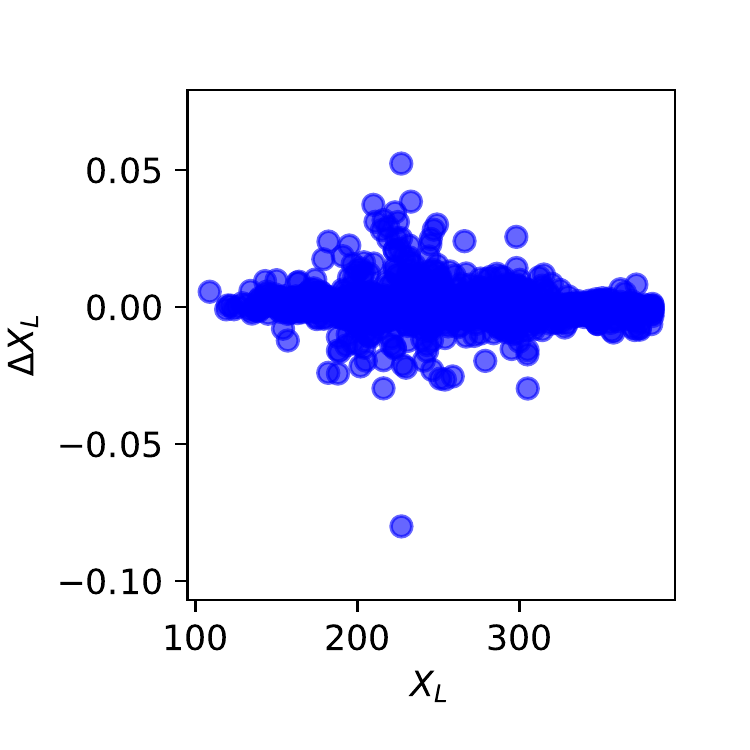}
\includegraphics[width=0.21\textwidth,height=0.13\textheight]{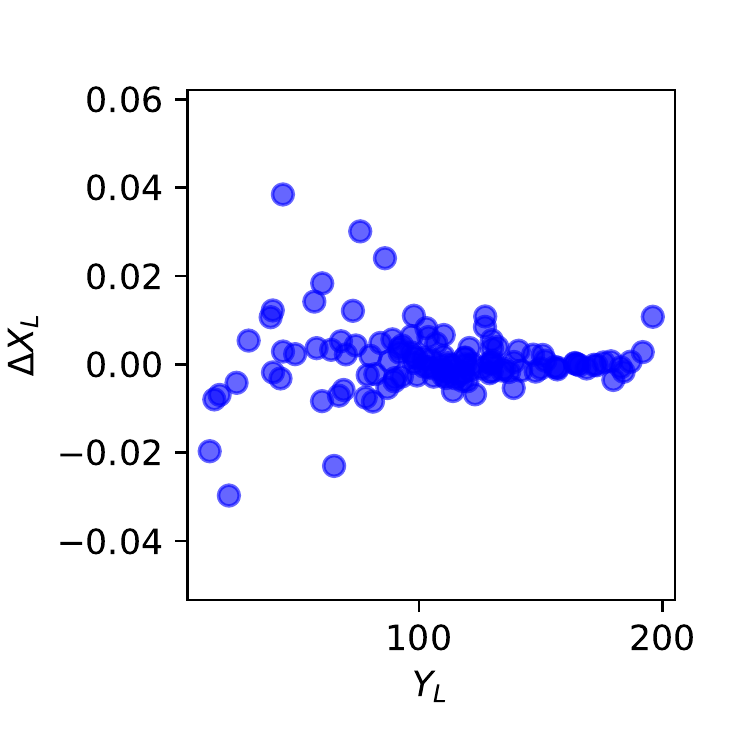}
\includegraphics[width=0.21\textwidth,height=0.13\textheight]{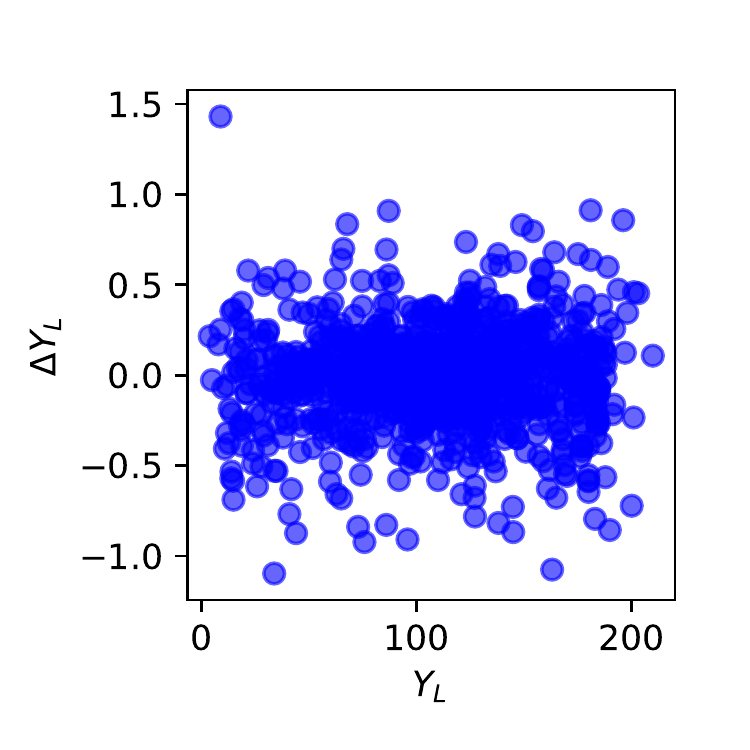}
\includegraphics[width=0.21\textwidth,height=0.13\textheight]{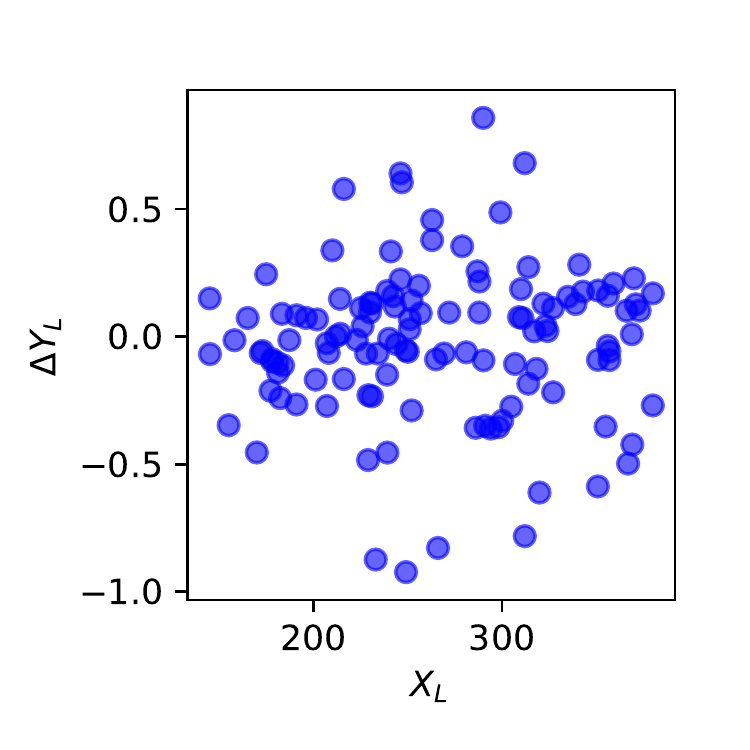}
\includegraphics[width=0.21\textwidth,height=0.13\textheight]{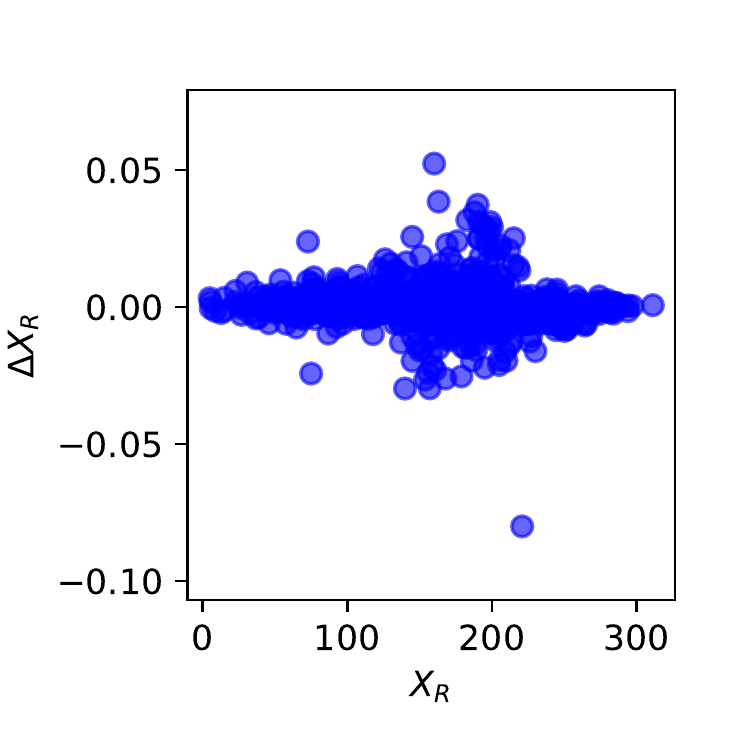}
\includegraphics[width=0.21\textwidth,height=0.13\textheight]{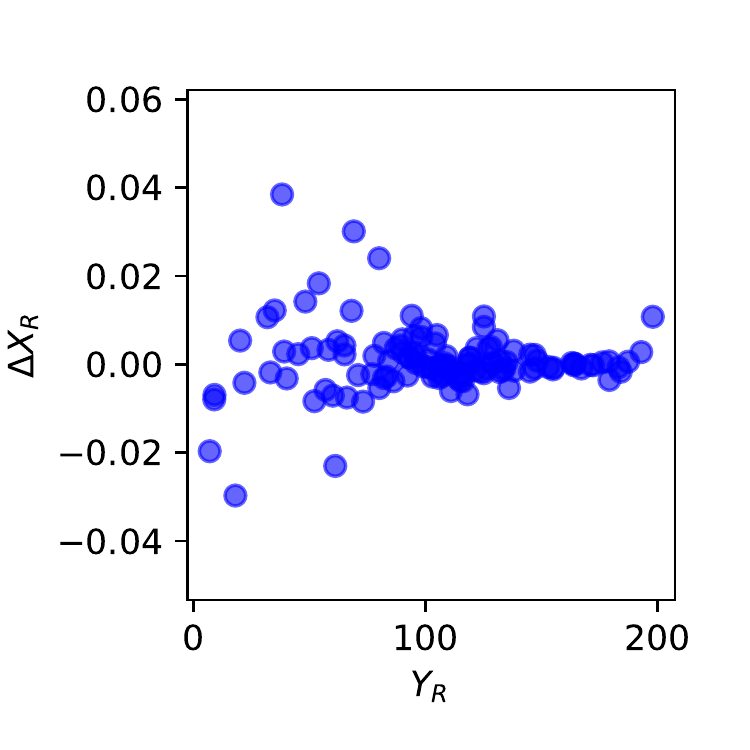}
\includegraphics[width=0.21\textwidth,height=0.13\textheight]{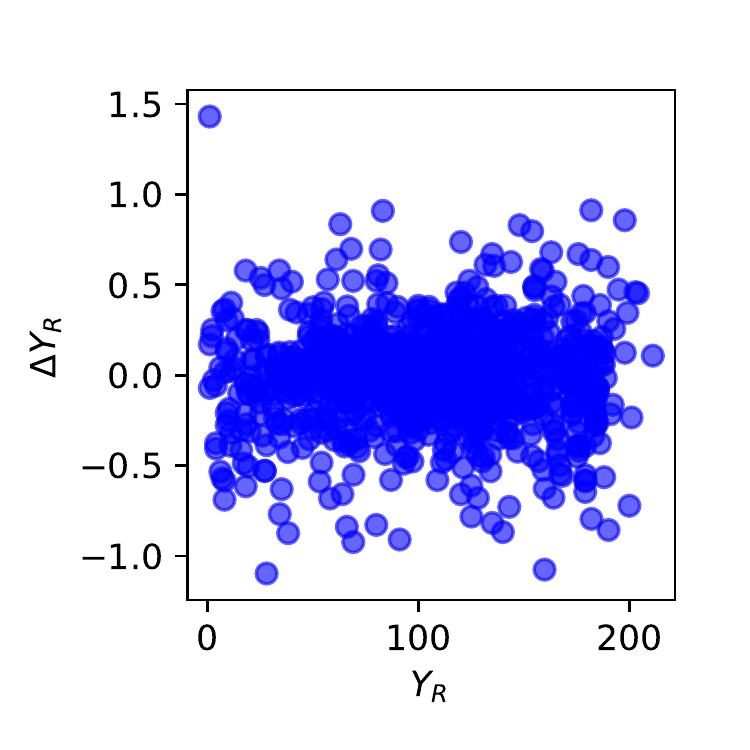}
\includegraphics[width=0.21\textwidth,height=0.13\textheight]{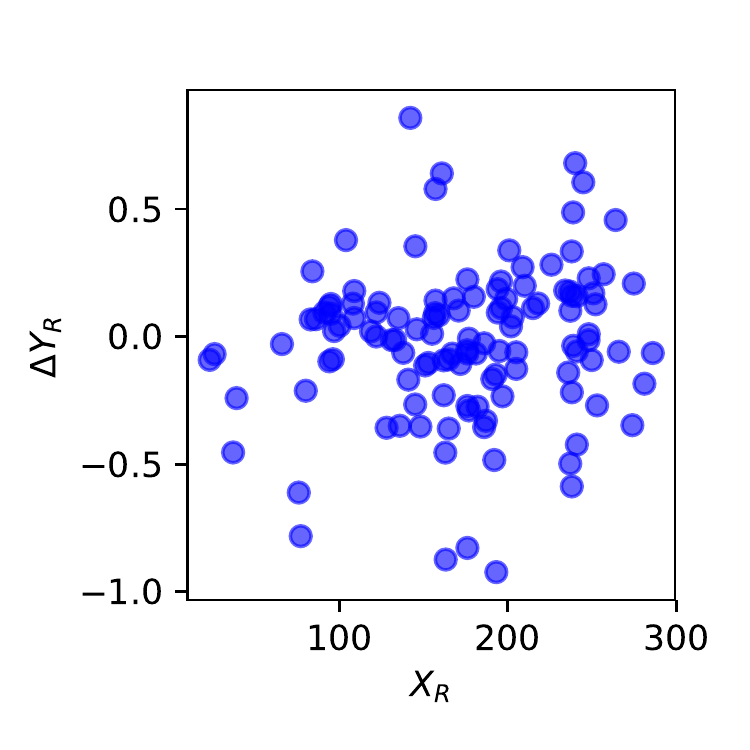}
\caption{Error field vs position. The deviation $\Delta x$ is widely distributed around the central region of the $x$-axis and the top region of the $y$-axis. The deviation $\Delta y$ is quite uniformly distributed along the $x$-axis and $y$-axis. }
\label{fig:errorfunc}
\end{figure}

For each point, the vector from this point to the closest epipolar line is defined as ``error field". Figure~\ref{fig:errorstat} shows the distributions of the error field of these 1000 pairs. The error field is found to have a mean deviation along the $x$-axis ($\Delta x$) within 0.01 pixels and along the $y$-axis ($\Delta y$) within 0.5 pixels. The mean uncertainty is larger along the $y$-axis. Figure~\ref{fig:errorfunc} shows the deviation at different positions of points in the image. It is quite uniform for $\Delta y$ but widely distributed for $\Delta x$ around the central region of the $x$-axis and the top region of the $y$-axis. The error needs to be considered for the 3D trajectory reconstructions. 

\section{3D Track Reconstruction}
\label{sec:3Dtracking}
\begin{figure}[t]
\includegraphics[width=0.23\textwidth]{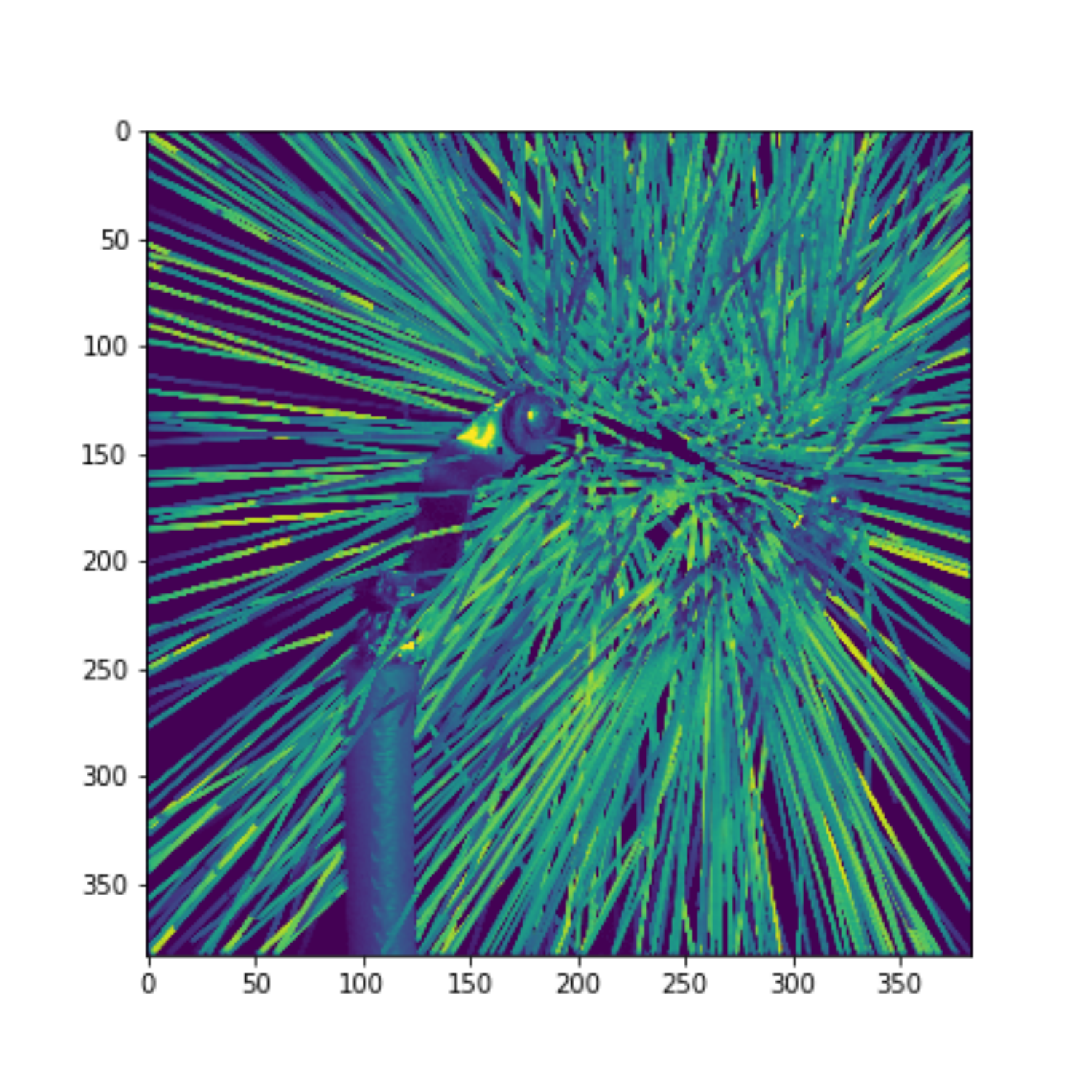}
\includegraphics[width=0.23\textwidth]{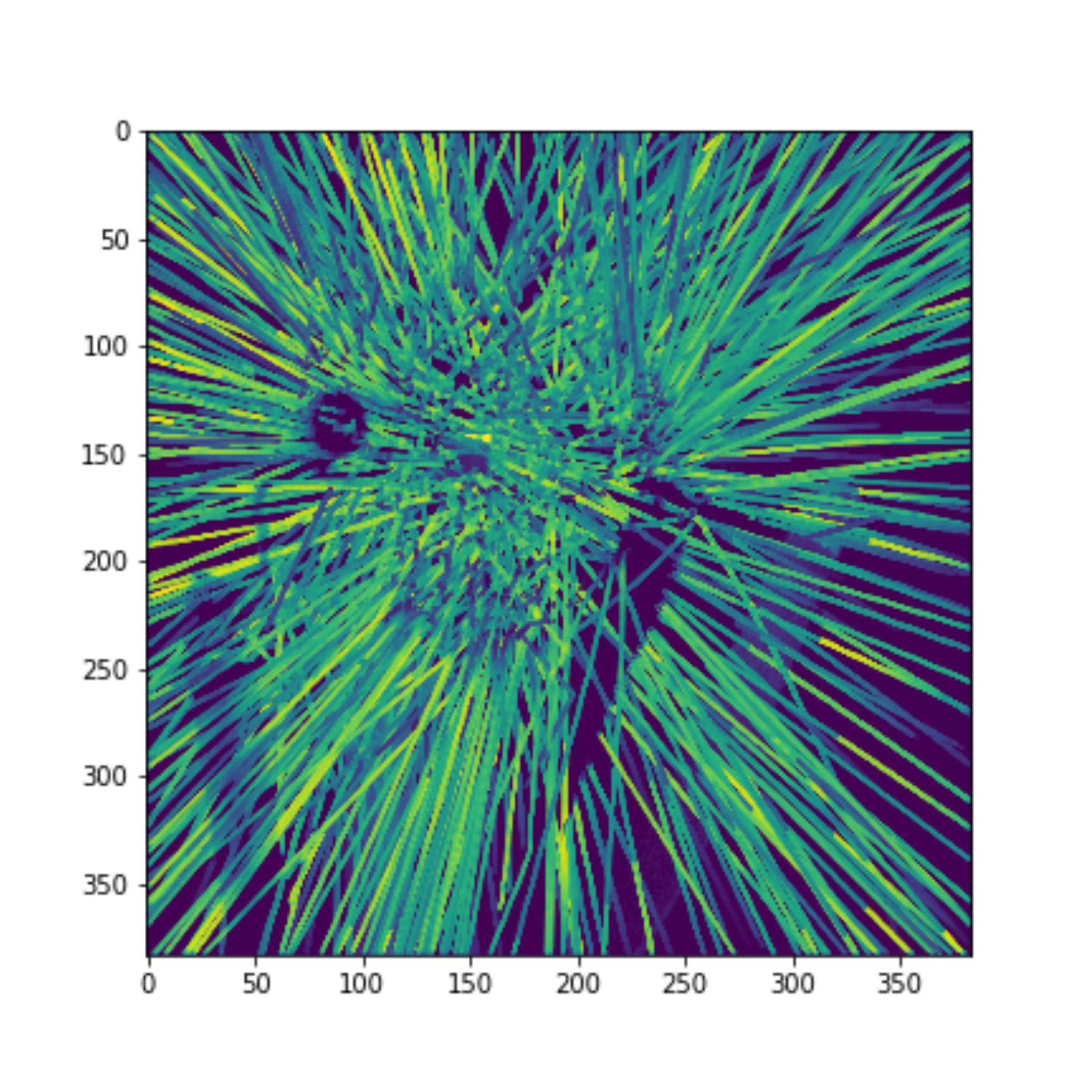}
\includegraphics[width=0.23\textwidth]{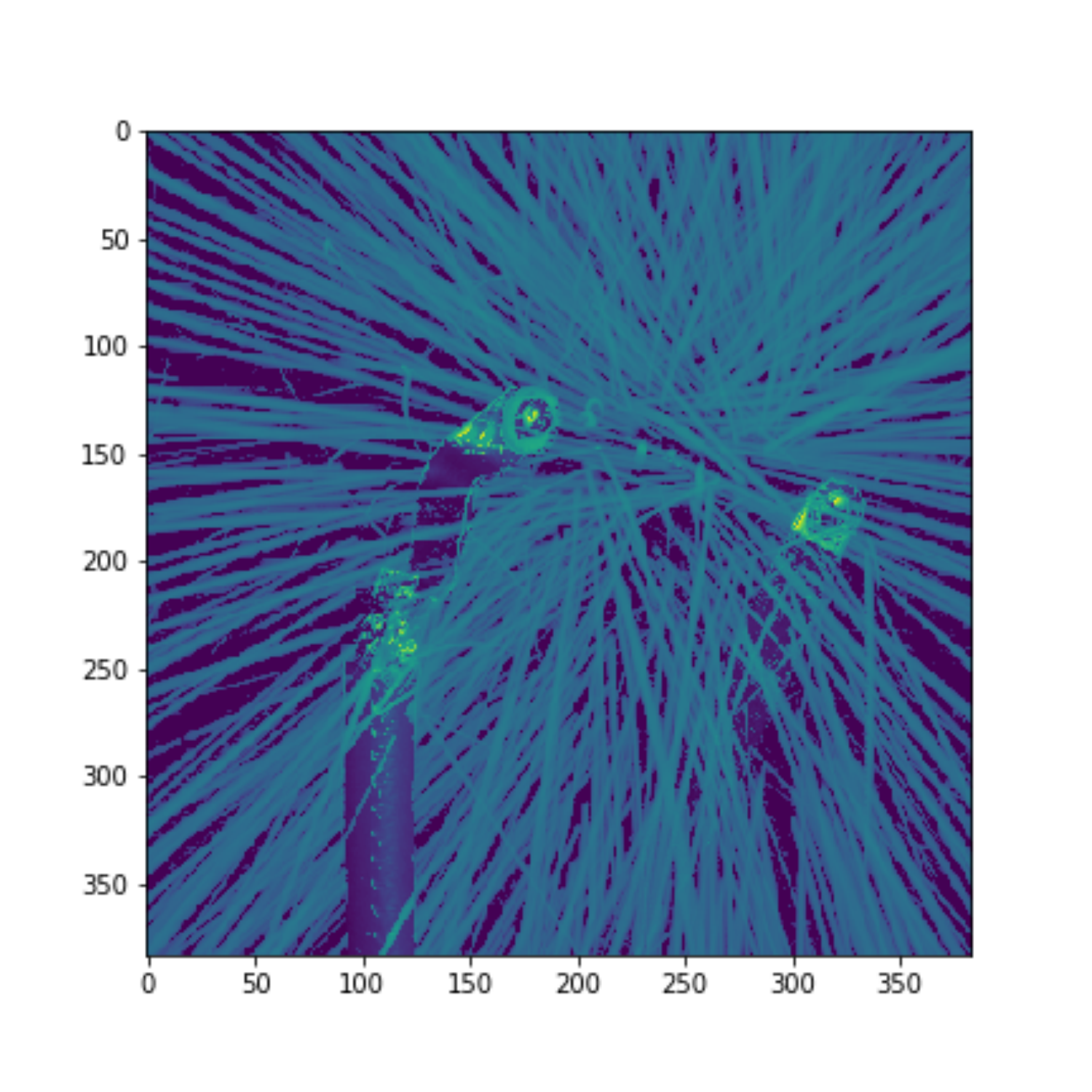}
\includegraphics[width=0.23\textwidth]{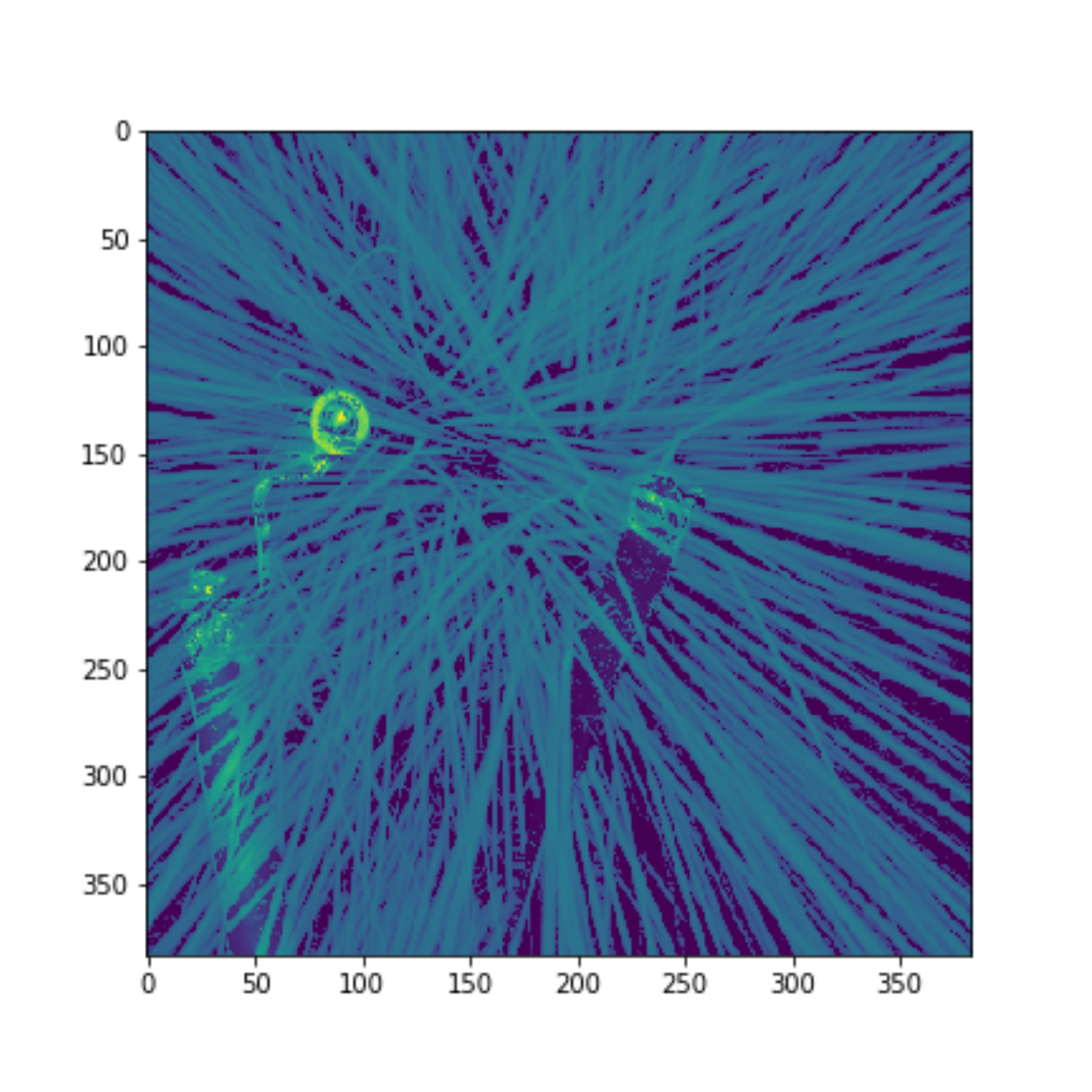}
\caption{Top two images are total tracks of Shot232 identified by \texttt{TrackPy}. Bottom two images are time-integrated images by superimposing the whole movie sequences. The similarity between the top and bottom images shows that \texttt{TrackPy} can identify most of tracks. Left (right) images correspond to the left (right) camera.}
\label{fig:alltracks_119}
\end{figure}

\texttt{TrackPy}~\cite{TrackPy}, an open-source python package, is used to identify 2D tracks from the movies by each camera. The \texttt{NearestVelocityPredict} function in the package is applied to link points in different frames. This function can estimate the velocity of the point in the current time frame and link it to the point with the closest velocity in the next time frame. We also reversed the raw movie sequences in the analysis so that the program can identify tracks starting from less crowded frames. However, some tracks are too short because of failure of linking. Some tracks are split because of noise or overlapping of tracks in the image. We applied two cuts to remove nonphysical tracks and saved tracks spanning more than 200 frames and of the maximum length larger than 20 pixels. Figure~\ref{fig:alltracks_119} shows that the majority of the tracks of Shot232 can be identified by \texttt{TrackPy} after the cuts where the color of tracks means the brightness identified by \texttt{TrackPy}. The similarity between the image of tracks and superimposition shows \texttt{TrackPy} is reliable for tracking in 2D. 

\begin{figure}[ht]
\includegraphics[width=0.45\textwidth]{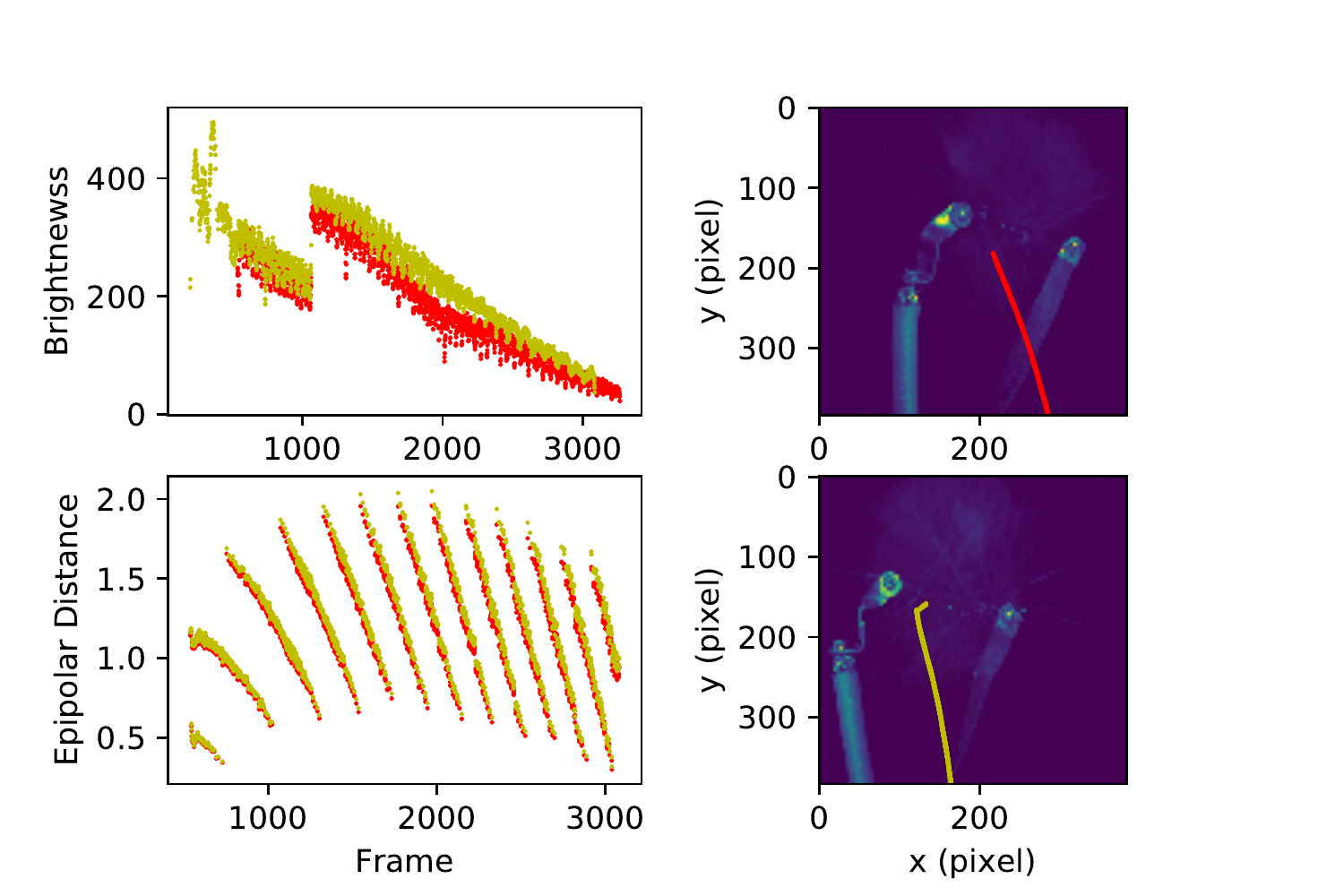}
\caption{One example of track pairing in Shot232. The red track is the one in the left camera and the yellow track is the one in the right camera. It shows their similar brightness and close epipolar distance. The structure of the epipolar distance seems to be related to the spatial resolution and calibration error.}
\label{fig:pairing}
\end{figure}

After finding the 2D tracks in each camera image, temporal evolution of track brightness and minimizing the epipolar distance can efficiently identify a large number of track pairs. Temporal evolution of track brightness can be used for track pairing from two views~\cite{Wang:2016}. Together with the epipolar distance defined in Sec.~\ref{sec:calibration}, we can efficiently identify a large number of track pairs. Figure~\ref{fig:pairing} shows one example of successful track pairing. The figures on the right show the tracks in the image of the left and right cameras, respectively. The left top figure shows the brightness evolution of the two tracks, which show similar temporal behaviors. The left bottom figure shows the epipolar distance calculating from the point pair of two tracks in corresponding frames. As described in Sec.~\ref{sec:calibration}, the epipolar distance is not always zero because of image noise or poor spatial resolution as well as the spatial calibration and the particle center calculation. Here the epipolar distance of the paired track pair is no more than 2 pixels. A good pair must have small epipolar distances. A computer program is used to calculate epipolar distances for random track pairs from the left and right cameras. The results led to reduced number of possible pairing based on similar brightness evolution and small epipolar distances. Final pairings were selected manually here but they could be automated later. The analysis code is stored in Ref.~\cite{code}.

\begin{figure}[ht]
\includegraphics[width=0.5\textwidth]{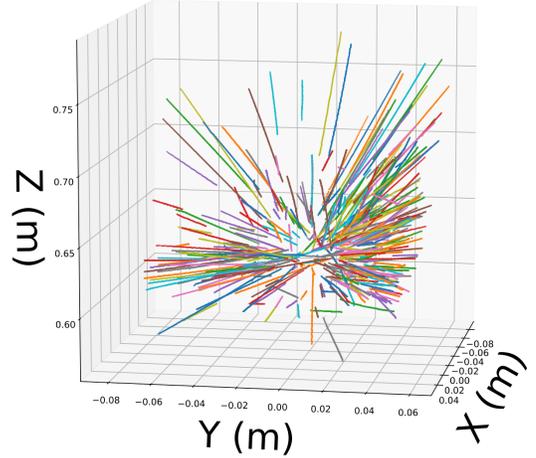}
\caption{All reconstructed 3D tracks which are exploding from the central region to outside. }
\label{fig:3Dtracks}
\end{figure}

Following the track-pairing from the two views, triangulation is used to reconstruct 3D tracks~\cite{Wang:2016}. Figure~\ref{fig:3Dtracks} summarizes all reconstructed 3D tracks in this work. Tracks convergence near the center is consistent with the wire location before the explosion. Particle motion along each track can be analyzed using various models.

\section{Preliminary Force Analysis}

\label{sec:force}
\begin{figure}[ht]
\includegraphics[width=0.5\textwidth,height=0.33\textheight]{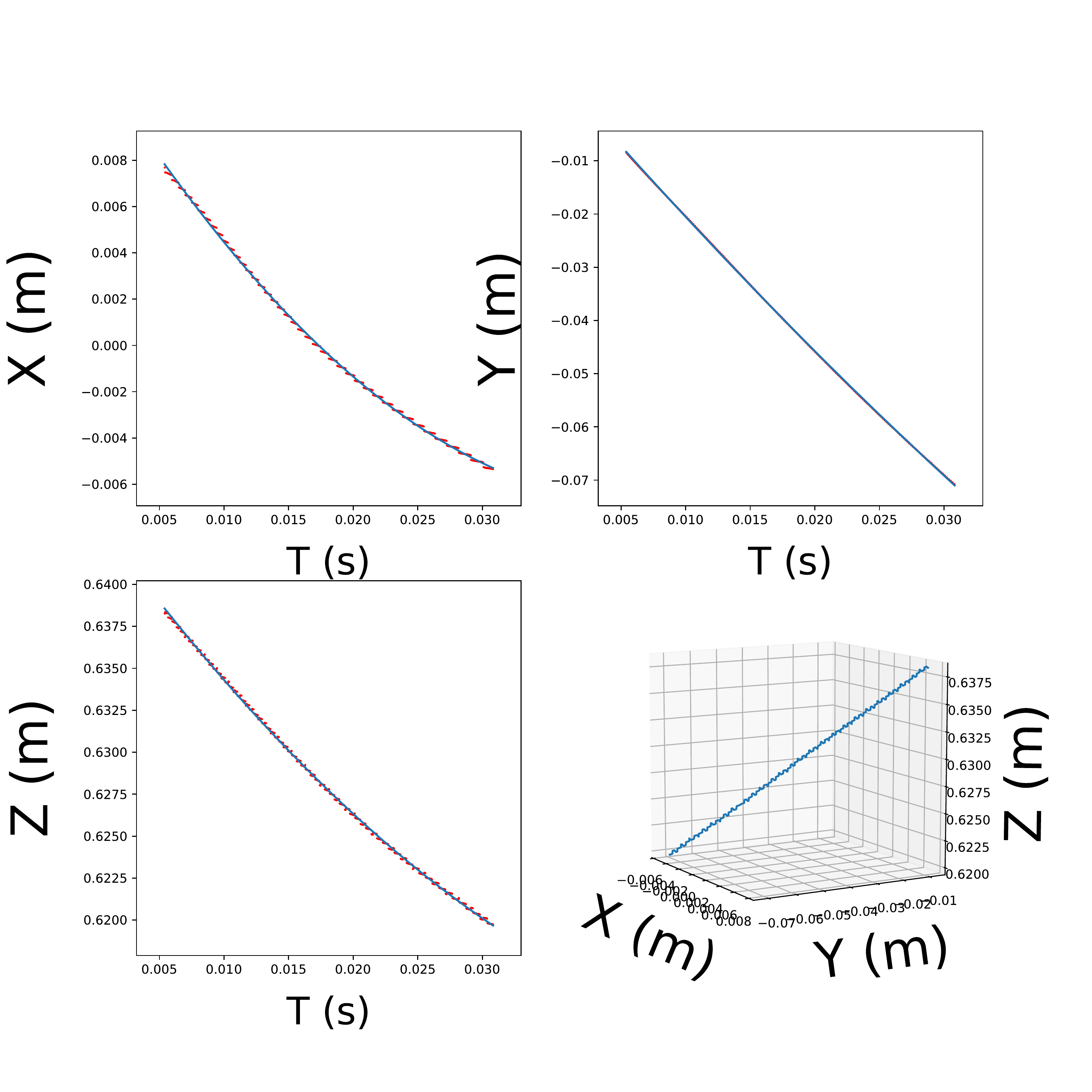}
\caption{One example of 3D fitting in Shot232. The fitting parameters are $X(T) = 0.0123-0.8924T+10.3828T^2$, $Y(T)=0.0065-2.7988T+9.1978T^2$ and $Z(T) =0.6440-1.0617T+8.7934T^2$.}
\label{fig:3Dfitting}
\end{figure}
\begin{figure}[ht]
\includegraphics[width=0.5\textwidth]{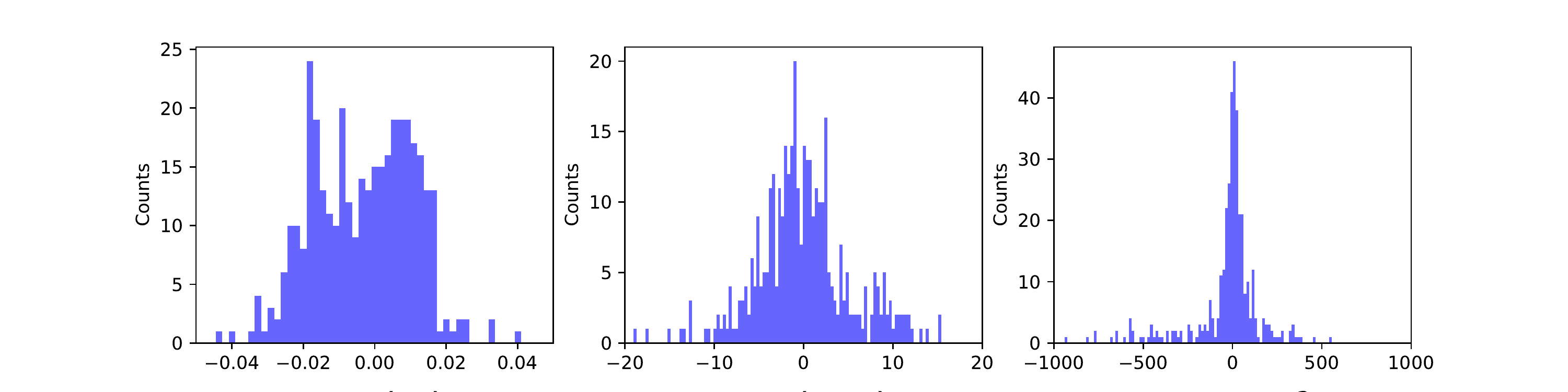}
\includegraphics[width=0.5\textwidth]{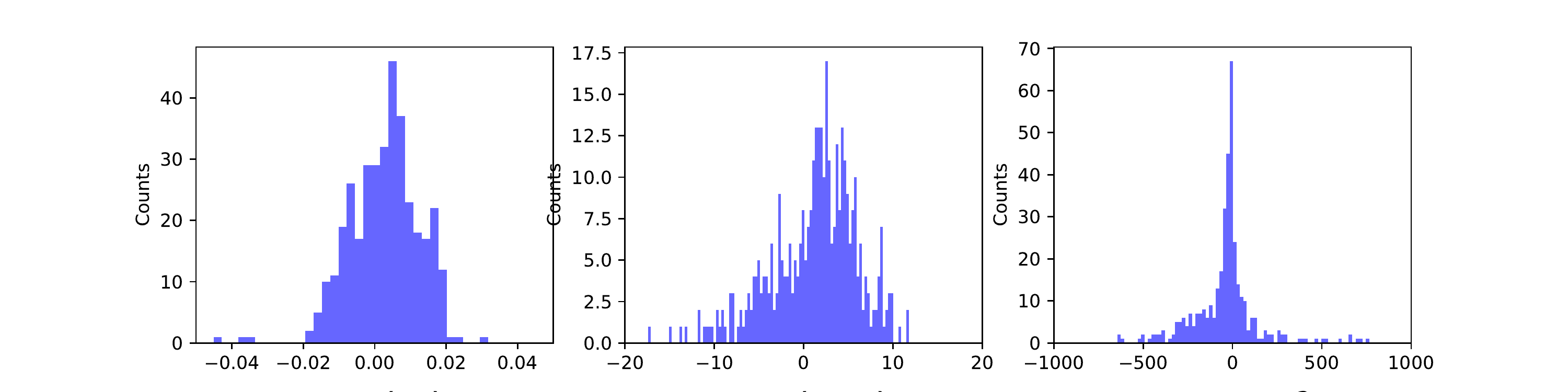}
\includegraphics[width=0.5\textwidth]{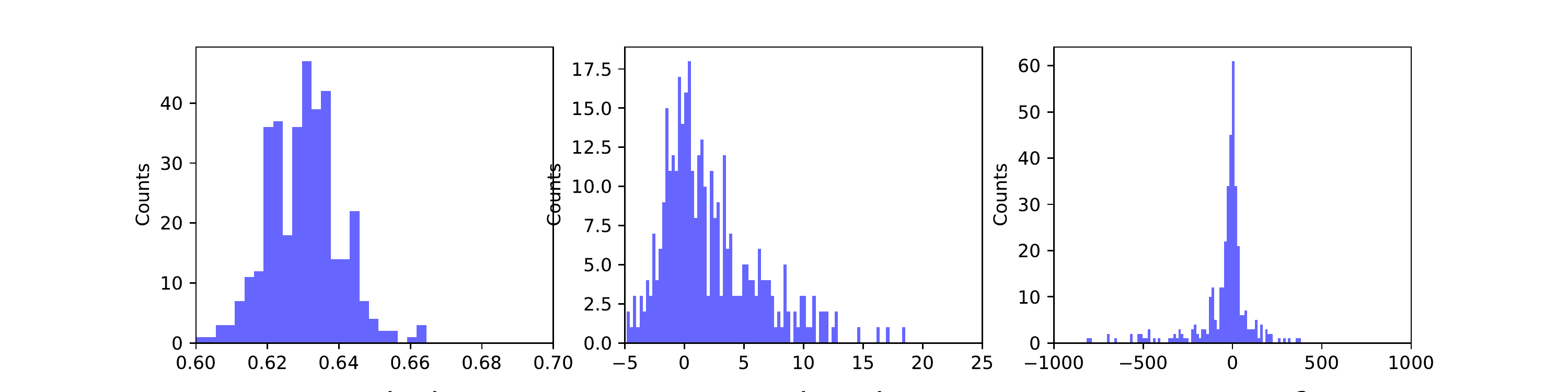}
\caption{Fitting parameters from top to bottom)$x$-axis, $y$-axis and $z$-axis; from left to right) $P_0$, $V$ and $A/2$. It shows the initial position of most tracks is around $x\sim 0, y\sim 0$ and $z\sim 0.63$.} \label{fig:acceleration_XYZ}
\end{figure}

We may use a parabolic function of time ($T$) to fit the coordinates of different particle trajectories, 
\begin{align}
    f(T) = P_0 + VT + A/2T^2
\end{align}
where $P_0$ is the initial position at $T=0$, $V$ is the initial velocity and $A$ is the  acceleration  displayed as vector quantities. Figure~\ref{fig:3Dfitting} compares the 3D reconstruction of the track in Fig.~\ref{fig:pairing} with a parabolic curve fitting along each axis. As shown in the figure, the track has steps along each axis because of finite pixel size. This pixel size causes poor spatial resolution or smearing due to finite exposure time. However, we can fit tracks using parabolic curve to derive velocity and acceleration information. Figure~\ref{fig:acceleration_XYZ} shows the population of parameters of all data. Initial offsets of the $x$-axis and $y$-axis are centered around the origin while the offset of the $z$-axis is around 0.63 m. %The total velocity is equal to  

\begin{figure}[ht]
\includegraphics[width=0.5\textwidth]{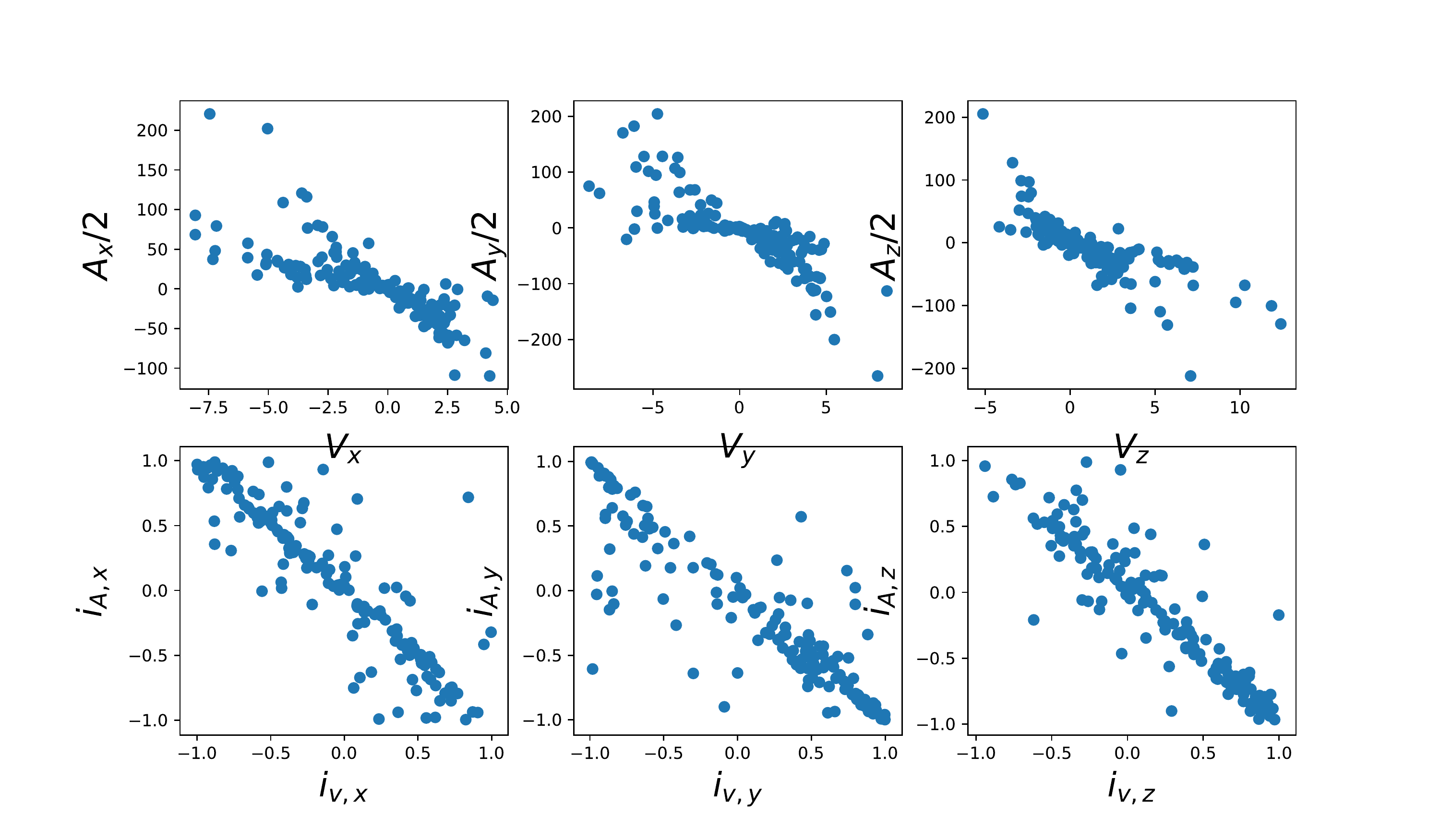}
\caption{Top) Acceleration vs. velocity. Bottom) Acceleration direction vs. velocity direction. It implies the acceleration is proportional to velocity. Different slopes may be caused by different particle size in different shots.}
\label{fig:acceleration_velocity}
\end{figure}

If we ignore the `outliers' in the fitted acceleration, {\it i.e.}, the fitting uncertainty of the acceleration from the main population being larger than 2, Figure~\ref{fig:acceleration_velocity} shows the remaining distributions of acceleration and velocity. It is found that the acceleration is proportional to the velocity. The most probable acceleration is less than 100 m/sec$^2$, or $\sim$ 10 g with $g$ being the gravitational acceleration. The preliminary result suggests the acceleration is proportional to the velocity, implying the existence of viscosity. We do not expect any electromagnetic forces for the experiment for several reasons: no charge expected; no significant electric or magnetic field exist in the experiment. That leaves the viscous force due to the ambient air as the best force candidate other than gravity. Additional work will be pursued in the future.

\section{Conclusion}
\label{sec:conclusion}
As a part of high-speed imaging tracking diagnostics for high-temperature microparticles, we have implemented a new particle tracking algorithm and demonstrated particle tracking using two fast cameras in-sync. This synthetic diagnostics with two fast cameras in-sync provides a robust method for detecting dust effects in plasmas. Force analysis indicates that the acceleration on the order of 10 g can be measured. Viscous force may be important in addition to gravity in particle motion for these experiments, which were performed in the ambient air environment. The method will be used in plasma experiments to study plasma-particle interactions and other magnetic fusion scenarios.

\section{ACKNOWLEDGMENTS}
This work is supported by the U.S. Department of Energy/Fusion Energy Sciences program.
\bibliography{main}
\end{document}